\documentclass[11pt]{article}

\usepackage[margin=1in]{geometry}
\usepackage{times}
\usepackage{amsmath,amssymb}
\usepackage{graphicx}
\usepackage{booktabs}
\usepackage{multirow}
\usepackage{hyperref}
\usepackage{xcolor}
\usepackage{listings}
\usepackage{caption}
\usepackage{enumitem}
\usepackage{authblk}

\hypersetup{
    colorlinks=true,
    linkcolor=blue,
    citecolor=blue,
    urlcolor=blue
}

\definecolor{codebg}{rgb}{0.96,0.96,0.96}
\title{Fine-Tuning Qwen3-27B for C-to-Rust Code Translation: \\
A Three-Stage Curriculum of Pretraining, Debugging-Aware SFT, and Task-Specific SFT}

\author[1]{Pu Zhao}  \author[1]{Changdi Yang} \author[1]{Yixiao Chen}  \author[1]{Yi Gao} 
\author[2]{Yifan Cao} \author[3]{\\Haochen Zeng} \author[1]{Yanzhi Wang}
\affil[1]{Northeastern University}
\affil[2]{EmbodyX Inc}
\affil[3]{Aibao LLC}

\begin{document}
\date{}
\maketitle

\begin{abstract}
Translating C code into safe, idiomatic Rust is a longstanding software-engineering goal because it can eliminate entire classes of memory-safety vulnerabilities while preserving the functional behavior of legacy systems. Large language models (LLMs) have shown promise for this task but typically underperform when applied off-the-shelf, since general-purpose pretraining rarely emphasizes idiomatic Rust generation, cross-language semantic equivalence, or the ability to reason about and repair compiler/runtime feedback. In this report we describe a three-stage fine-tuning curriculum applied to Qwen3-27B that is designed to progressively specialize the model for the C-to-Rust (C2Rust) translation task: (1) continued pretraining on Rust-centric corpora to strengthen the model's prior over idiomatic Rust syntax and standard-library usage; (2) supervised fine-tuning (SFT) on the \texttt{microsoft/Verus\_Training\_Data} dataset to instill debugging and self-repair behavior over Rust code; and (3) task-specific SFT on paired C/Rust solutions derived from LeetCode problems to teach direct semantic translation. We evaluate the resulting model using the agentic, static-analysis-guided verification framework of SACTOR, which performs structure-aware, two-phase (unidiomatic $\to$ idiomatic) translation with foreign-function-interface (FFI)-based end-to-end (E2E) testing. We report success rate, idiomaticity (Clippy lint counts, unsafe-code fraction), and failure-mode analyses, and compare our fine-tuned model against baseline Qwen3-27B and other LLMs evaluated under the same framework. \\
Model: \textit{https://huggingface.co/moxin-org/C2Rust}\\
Code: \textit{https://github.com/moxin-org/C2Rust} 
\end{abstract}

\tableofcontents

\section{Introduction}
\label{sec:intro}

C remains the implementation language of choice for a large fraction of performance-critical and systems-level software, but its manual memory management is a persistent source of vulnerabilities such as buffer overflows, use-after-free errors, and memory leaks. Rust addresses these issues through a compile-time ownership and borrowing model that guarantees memory safety without a garbage collector, and it has seen growing adoption in security-sensitive contexts (e.g., the Linux kernel, Firefox). Migrating existing C codebases to Rust therefore has clear practical value, but manual translation is slow, expensive, and error-prone, motivating automated and LLM-assisted translation approaches.

Rule-based tools such as C2Rust~\cite{immunant2020c2rust} generate syntactically valid Rust by transliterating C ASTs, but the resulting code is heavily unsafe and unidiomatic. Recent work has explored LLM-driven translation with various forms of verification and feedback (e.g., fuzzing, symbolic execution, FFI-based E2E testing) to improve both correctness and idiomaticity~\cite{sactor2025}. However, general-purpose LLMs \cite{grattafiori2024llama,team2026qwen3,shen2024search,mi2026effective,shen2025sparse,zhan-etal-2024-rethinking-token,zhao-etal-2024-pruning,zhao2025open,zhao20247b} are not natively specialized for this task: they may lack deep familiarity with idiomatic Rust patterns, may not reliably reason about ownership/lifetime semantics, and may struggle to use compiler and test feedback to iteratively repair their own output.

The C2Rust setting we study is best situated within the much broader trajectory of LLMs applied to code more generally. Autoregressive, decoder-only language models trained on very large source-code corpora have become the dominant approach to automated program synthesis, translation, and repair, and progress in that broader literature has come from three largely complementary directions. The first is simply scaling: successive generations of code-oriented LLMs have grown in both pretraining-corpus size and parameter count, and general software-engineering benchmarks such as SWE-bench Verified~\cite{jimenez2024swebench} (which we also use in \S\ref{sec:swebench}) have tracked steady gains from this scaling alone. The second is post-training: instruction tuning and, increasingly, reinforcement learning with verifiable rewards (i.e., using compiler or test-execution outcomes directly as a training signal) have proven effective at aligning a fixed-capacity model's outputs with the kind of structured, executable feedback that code tasks naturally provide. The third is inference-time scaffolding: agentic tool use, retrieval of relevant context, and iterative repair loops let a model make repeated, feedback-conditioned generation attempts rather than committing to a single one-shot output. Our own three-stage curriculum  and our adoption of SACTOR's agentic, repair-driven evaluation loop  are best understood as instances of, respectively, the second and third of these directions, specialized to the narrower C-to-Rust translation task rather than as departures from the broader LLM-for-code literature.

A more recent and, so far, largely separate line of work explores diffusion-based language models (dLLMs) as an alternative generative paradigm to autoregressive decoding for code. Rather than producing tokens strictly left-to-right, masked or discrete diffusion models iteratively denoise an entire sequence, refining many token positions in parallel across a fixed number of denoising steps; this global, non-causal generation process has been motivated for code specifically on the grounds that writing code often involves non-sequential, back-and-forth revision rather than strict left-to-right composition. Open-weight dLLMs such as LLaDA~\cite{nie2025llada} and Dream~\cite{ye2025dream}, together with code-specialized variants such as DiffuCoder~\cite{gong2025diffucoder} and Dream-Coder~\cite{xie2025dreamcoder}, have shown that diffusion-based decoding can approach the quality of similarly sized autoregressive models on coding benchmarks, while commercial-scale systems such as Mercury~\cite{inceptionlabs2025mercury} and Gemini Diffusion~\cite{deepmind2025geminidiffusion} have demonstrated diffusion-based code generation that is competitive with leading autoregressive code models while offering substantially faster generation. At the same time, the dLLM ecosystem remains considerably less mature than that of autoregressive LLMs: serving optimizations that are now standard for autoregressive decoding, such as key-value caching, speculative decoding, and prefix caching, have no direct analogue in the diffusion setting, and dLLMs typically must commit to an output length in advance rather than terminating generation naturally via an end-of-sequence token. We view diffusion-based code models as a plausible complementary direction for future C2Rust work rather than as directly applicable to the present study: notably, the idiomatic-refinement phase of the SACTOR pipeline we adopt  is itself a form of iterative, whole-program revision rather than strictly incremental generation, a structural resemblance to denoising that could make dLLMs a natural fit for that specific sub-task, though we leave an empirical investigation of this possibility to future work. Throughout this report, our base model and all three curriculum stages retain the standard autoregressive, next-token-prediction formulation, and we do not otherwise revisit diffusion-based decoding below.

This report documents an effort to specialize \textbf{Qwen3-27B} for the C2Rust translation task via a three-stage fine-tuning curriculum:

\begin{enumerate}[label=\textbf{Stage \arabic*:}, leftmargin=1.6cm]
    \item \textbf{Rust-focused continued pretraining.} We continue pretraining the base model on a corpus of Rust code and text collected from multiple existing Rust-capable models/sources, with the goal of strengthening the model's general command of Rust syntax, idioms, and standard-library usage prior to any task-specific tuning.
    \item \textbf{Debugging-oriented SFT.} We fine-tune on \texttt{microsoft/Verus\_Training\_Data}, an instruction dataset centered on debugging Rust code (e.g., diagnosing and repairing compiler errors, verification failures, or semantic bugs), with the aim of teaching the model to interpret error signals and produce targeted repairs -- a skill directly relevant to the iterative, feedback-driven verification loop used at evaluation time.
    \item \textbf{C2Rust task-specific SFT.} We fine-tune on instruction data constructed from LeetCode problems for which both a C solution and a corresponding Rust solution are available, directly teaching the source $\to$ target translation mapping the model must perform at inference time.
\end{enumerate}

We evaluate the final model using the agentic evaluation framework described by SACTOR~\cite{sactor2025}, which decomposes translation into an interface-preserving \emph{unidiomatic} phase and a refinement-based \emph{idiomatic} phase, each verified via FFI-linked E2E tests against the original C program, with structured compiler/test feedback driving up to $N$ repair attempts. Our model with 27B parameters demonstrates a success rate of 87.2\%, outperforming MiniMax-M2.5 and GLM-5  with hundreds of billions parameters. 

\subsection{Contributions}
\begin{itemize}
    \item A three-stage curriculum -- Rust pretraining $\to$ debugging SFT $\to$ C2Rust SFT -- for specializing a general-purpose 27B-parameter LLM to the C2Rust translation task.
    \item An evaluation of the resulting model under the SACTOR agentic verification framework, reporting success rate and idiomaticity metrics.

\end{itemize}

\section{Related Work}

\paragraph{Rule-based and static C-to-Rust translation.} C2Rust~\cite{immunant2020c2rust} performs AST-level transliteration from C to Rust, producing code that compiles but relies heavily on unsafe blocks and raw pointers, yielding low readability and limited safety benefit; Hong and Ryu's survey~\cite{hong2025automatically} traces the broader landscape of automatic C-to-Rust translation techniques that have followed. Subsequent static/rule-driven refinements target specific unidiomatic patterns left behind by transliteration -- for example, replacing C-style output parameters with algebraic-data-type return values, translating C's tagged unions into Rust's enum-based tagged unions, and automatically deriving lock APIs for concurrent C code -- but remain fundamentally rule-driven and struggle to generalize to complex, real-world code. Hong and Ryu also explore a hybrid point in this space, using an LLM specifically to resolve the type-migration step of C-to-Rust translation while keeping the surrounding pipeline rule-based~\cite{hong2025type}.

\paragraph{Single-shot and prompting-based LLM translation.} A first wave of LLM-driven approaches treats C-to-Rust translation largely as a prompting problem, optionally combined with post-hoc verification signals such as fuzzing, symbolic execution, or dynamic-analysis-guided repair. SafeTrans~\cite{farrukh2025safetrans} evaluates six LLMs across nearly 16,000 translation attempts, combining basic compiler-feedback repair with a few-shot guided repair phase keyed to specific Rust error categories (trait-implementation failures, borrow-checker violations), and additionally studies whether memory-safety vulnerabilities in the original C code persist after translation. C2SaferRust~\cite{nitin2025c2saferrust} combines LLM-driven translation with neurosymbolic techniques aimed at reducing residual unsafe code, though SACTOR's own analysis notes it still produces Rust with numerous unsafe blocks~\cite{sactor2025}. VERT~\cite{yang2025vert} targets verified-equivalent transpilation, using LLMs as few-shot learners to produce Rust translations whose behavioral equivalence to the source can be formally checked, though it is reported to struggle on programs with complex data structures~\cite{sactor2025}.

\paragraph{Verification- and feedback-guided translation.} Closest to our own evaluation setting, SACTOR~\cite{sactor2025} proposes a two-stage, structure-aware pipeline (interface-preserving unidiomatic translation, followed by idiomatic refinement) guided by static analysis and verified end-to-end via FFI linking with the original C program; a SPEC-driven harness generator bridges ABI mismatches introduced by idiomatic refactoring. We adopt this framework, unmodified, as our evaluation harness so that our fine-tuned model's performance can be compared against the LLMs benchmarked in that work (e.g., GPT-4o, Claude 3.5 Sonnet, Gemini 2.0 Flash, Llama 3.3 70B, DeepSeek-R1). AdaTrans~\cite{liu2026adatrans} similarly builds a generate-verify-repair loop around compiler feedback, but adapts its repair strategy to the specific category of failure observed (syntax vs.\ ownership vs.\ behavioral errors) via an error-stratified retrieval-augmented generation mechanism, reporting a 95.51\% compilation pass rate and 81.09\% solve rate on a 104-problem algorithmic benchmark; Syzygy~\cite{shetty2024syzygy} takes a related dual code-test translation approach, jointly translating a C program and its tests to Rust and using dynamic analysis to validate the pair.

\paragraph{Agentic and multi-step translation systems.} A further line of work frames C-to-Rust translation as a multi-step or agentic process rather than a single prompt-and-repair loop. Sim et al.~\cite{sim2025search} propose a search-based, multi-trajectory LLM agent that explores several candidate translation paths per program rather than committing to a single greedy repair sequence. AlphaTrans~\cite{ibrahimzada2025alphatrans} combines neural translation with symbolic validation at the repository level rather than per-function, aiming for compositional correctness across an entire codebase. Both are complementary to SACTOR's own two-phase design in that they explore alternative strategies for structuring the search over candidate translations under verifier feedback.

\paragraph{Project- and repository-level translation.} Because a single function or program is a limited proxy for real-world migration difficulty, several recent systems target project-scale translation directly. EvoC2Rust~\cite{wang2026evoc2rust} first generates a compilable Rust ``skeleton'' of type-checked function stubs for an entire project, then incrementally fills in each function and repairs remaining compilation errors via static analysis, avoiding a conventional transpiler-preprocessing step. His2Trans~\cite{wang2026his2trans} extends this skeleton-first idea with a self-evolving historical-retrieval mechanism, mining previously completed migrations to infer idiomatic API correspondences and reduce repeated repair overhead on unseen projects by an order of magnitude. PtrTrans~\cite{yuan2025project} instead constructs a C-Rust pointer knowledge graph encoding ownership, mutability, nullability, and lifetime information extracted via program analysis, injecting these facts into LLM prompts to guide project-level, ownership-compliant generation. These project-level systems are the closest existing analogs to the \texttt{projects/} granularity of the C2Rust-Moxin dataset described in \S\ref{sec:stage3} (though, as noted there, our Stage 3 SFT itself draws only on the \texttt{functions/} and \texttt{programs/} granularities).

\paragraph{Benchmarks and empirical studies.} Beyond CRust-Bench~\cite{khatry2025} and the TransCoder-IR~\cite{szafraniec2023}/CodeNet~\cite{puri2021} datasets used in the SACTOR evaluation protocol, several other benchmarks and studies characterize LLM translation quality more broadly. CodeTransOcean~\cite{yan2023codetransocean} provides a large multilingual benchmark spanning many source/target language pairs (not limited to C/Rust), useful for situating C-to-Rust performance within the broader landscape of LLM code translation capability. Ou et al.~\cite{ou2024repobench} introduce a benchmark specifically targeting repository-level translation into Rust, complementing CRust-Bench's function-level focus. Li et al.~\cite{li2024translating} report a user study of practitioners translating C to Rust, providing qualitative grounding for what translation errors and idioms matter most to human developers -- a useful check against benchmarks that measure only automated pass/fail outcomes.

\paragraph{Instruction tuning and debugging-aware training.} Instruction fine-tuning on debugging/repair data has been used elsewhere to improve models' ability to consume compiler diagnostics and test failures and to produce targeted patches; the error-stratified repair strategies used by AdaTrans~\cite{liu2026adatrans} and the historical-repair-pattern reuse in His2Trans~\cite{wang2026his2trans} (discussed above) are inference-time analogs of the training-time debugging skill we aim to instill in Stage 2. We adopt microsoft/Verus Training Data~\cite{verusyn2026}, a Rust-centric formal-proof-debugging dataset (\S\ref{sec:stage2}), as an intermediate-stage curriculum designed to transfer this self-repair capability to the C2Rust setting, where iterative feedback (from the idiomatic/unidiomatic verifiers) is central to the evaluation protocol.

\paragraph{General-purpose LLMs for code, beyond C-to-Rust.} The C2Rust-specific literature surveyed above sits inside a much larger body of work applying LLMs to code more broadly, and several trends from that literature recur in our own design choices. Base-model scaling and code-specific continued pretraining -- the same strategy we use in Stage 1 (\S\ref{sec:stage1}) -- underlie essentially every strong general-purpose code model, from early code-pretrained encoder-decoder and decoder-only models through to today's frontier systems; StarCoderData~\cite{starcoderdata2023}, one of the seven Stage 1 sources, is itself the pretraining corpus behind an entire family of open code models rather than a C2Rust-specific artifact. Reinforcement learning with executable feedback -- compiler success, test-suite pass rates, or, in the case of Verus, formal-verifier acceptance -- has similarly become a standard post-training ingredient for code models generally, not only for translation systems; the VeruSyn pipeline underlying our Stage 2 data (\S\ref{sec:stage2}) is one instance of this broader pattern, in which a model is trained directly against a formal or executable success signal rather than against human preference labels alone. Finally, general software-engineering benchmarks such as SWE-bench~\cite{jimenez2024swebench} were developed specifically to measure this broader, task-agnostic notion of coding competence (issue localization, multi-file editing, real repository context) as a check against benchmarks -- including our own C2Rust evaluation  -- that isolate a single, narrower translation task; we report SWE-bench Verified results for exactly this reason in \S\ref{sec:swebench}.

\paragraph{Diffusion language models for code generation.} A separate line of work replaces autoregressive, left-to-right decoding with masked or discrete diffusion, in which a model iteratively denoises an entire token sequence rather than committing to tokens strictly in order \cite{shen2024lazydit,zhan2024fast,shen2025draftattention,shen2025fastcar,shen2025efficient,taherin2025cross,shen2025quartdepth,shen2024numerical,kong2025}. CodeFusion~\cite{singh2023codefusion} was the first system to combine a diffusion objective with code generation, but was restricted to small-scale models and comparatively simple synthesis tasks. More recent open-weight diffusion large language models (dLLMs), including LLaDA~\cite{nie2025llada} and its preference-optimized successor LLaDA~1.5~\cite{gu2025llada15}, and Dream~\cite{ye2025dream}, have scaled the masked-diffusion paradigm to parameter counts and benchmark performance competitive with similarly sized autoregressive LLMs; code-specialized variants such as DiffuCoder~\cite{gong2025diffucoder} and Dream-Coder~\cite{xie2025dreamcoder} adapt this paradigm specifically to large code corpora, and both report analyses of how a dLLM's denoising order and effective causality differ from autoregressive generation. At commercial scale, Mercury~\cite{inceptionlabs2025mercury} and Gemini Diffusion~\cite{deepmind2025geminidiffusion} demonstrate that diffusion-based code generation can match leading autoregressive code models on standard coding benchmarks while substantially reducing generation latency, since many token positions can be refined in parallel per denoising step rather than one token at a time. Li et al.~\cite{li2025survey} survey this broader emerging area, cataloguing architectural variants, training objectives, and open efficiency challenges across text and code applications of diffusion language modeling. We view this literature as a plausible source of future C2Rust methodology -- particularly for SACTOR's idiomatic-refinement phase, which is itself a whole-program revision step -- but the present work retains the standard autoregressive formulation throughout, and none of the training or evaluation described below involves a diffusion objective.

\section{Base Model}
\label{sec:base-model}
We use \textbf{Qwen/Qwen3.5-27B}~\cite{qwen35_2026} as the source model for our training. This checkpoint is part of the Qwen3.5 family released by the Qwen Team (Alibaba) in February 2026, and is distinct from the original Qwen3 dense lineup (0.6B/1.7B/4B/8B/14B/32B) released earlier: Qwen3.5 introduces a new hybrid-attention architecture and, in its base release, unified text/vision-language training. Qwen3.5-27B is released under the Apache 2.0 license.

\subsection{Architecture}
Qwen3.5-27B is natively a causal language model with an attached vision encoder (i.e., a unified multimodal checkpoint), but since our task (C2Rust translation) is purely textual, we use it in text-only / language-model-only mode throughout fine-tuning and inference, bypassing the vision tower. Table~\ref{tab:qwen35-arch} summarizes the language-model architecture as specified in the official model card.

\begin{table}[h]
\centering
\caption{Qwen3.5-27B language-model architecture (as released).}
\label{tab:qwen35-arch}
\begin{tabular}{@{}ll@{}}
\toprule
Property & Value \\
\midrule
Parameters (language model) & 27B ($\sim$28B including the vision encoder) \\
Hidden dimension & 5{,}120 \\
Number of layers & 64 \\
Layer layout & 16 $\times$ [\,3 $\times$ (Gated DeltaNet $\to$ FFN) $\to$ 1 $\times$ (Gated Attention $\to$ FFN)\,] \\
Gated DeltaNet heads & 48 (value), 16 (query/key) \\
Gated DeltaNet head dimension & 128 \\
Gated Attention heads & 24 (query), 4 (key/value) \\
Gated Attention head dimension & 256 \\
RoPE dimension (Gated Attention) & 64 \\
FFN intermediate dimension & 17{,}408 \\
Token embedding / LM head size & 248{,}320 (padded) \\
Multi-token prediction (MTP) & Trained with multi-step MTP head \\
Native context length & 262{,}144 tokens (YaRN-extensible to 1{,}010{,}000) \\
License & Apache 2.0 \\
\bottomrule
\end{tabular}
\end{table}

Architecturally, Qwen3.5-27B is therefore \emph{not} a conventional GQA-only Transformer: three out of every four layers use \textbf{Gated DeltaNet}, a gated linear-attention mechanism with a recurrent (constant per-step cost) formulation, while the fourth layer in each 4-layer block uses standard \textbf{Gated Attention} with grouped-query attention (24 query heads sharing 4 key/value heads) and rotary position embeddings. This hybrid design is intended to give near-linear-attention efficiency at long context lengths while retaining full attention's precision at fixed intervals -- relevant to our setting since C2Rust translation inputs (a full C source file, possibly with multiple functions/structs) plus SACTOR's iterative compiler/test feedback can produce long contexts within a single agent turn.

Fine-tuning  is performed directly on the released BF16 weights of \texttt{Qwen/Qwen3.5-27B}, with no architectural modifications: the model's layer layout, Gated DeltaNet/Gated Attention structure, hidden/FFN dimensions, and vocabulary are used unchanged from the official checkpoint (Table~\ref{tab:qwen35-arch}). Although Qwen3.5-27B is natively a unified vision-language checkpoint, all of our training data (Stages 1--3) is text-only -- no image or video inputs are used at any stage -- so the vision encoder is exercised only insofar as it remains part of the loaded checkpoint; it receives no gradient signal from our text-only instruction/pretraining data and plays no role in the C2Rust task itself. The vision-encoder weights were not stripped from the checkpoint prior to training; they were left in place but unused, since no image or video data appears anywhere in Stages 1--3. As a result, the full $\sim$28B-parameter checkpoint (language model plus vision encoder) is loaded and occupies memory during fine-tuning, even though only the $\sim$27B-parameter language-model component receives gradient updates from our text-only training data.

\subsection{Tokenizer and Licensing}
Qwen3.5-27B uses a padded vocabulary of 248,320 tokens (tied or separately-sized token embedding and LM head, per Table~\ref{tab:qwen35-arch}). The tokenizer and vocabulary are used unmodified throughout all three training stages -- no new special tokens (e.g., for delimiting C/Rust code blocks or instruction fields) were added; instruction formatting relies entirely on the model's existing chat template and vocabulary. The model is released under the Apache 2.0 license, which permits fine-tuning and redistribution of derivative weights, and no additional Qwen usage-policy terms apply beyond this license for the checkpoint used.

\subsection{Fine-Tuning Hardware}
All three training stages (\S\ref{sec:stage1}--\S\ref{sec:stage3}) were run on a single node of {8 NVIDIA B300 (Blackwell Ultra) GPUs}. 
Fine-tuning is performed in BF16, matching the precision of the released Qwen3.5-27B weights (no FP8/NVFP4 quantization or mixed-precision downcasting is applied during training, though NVFP4 remains available on Blackwell Ultra should later inference-time serving benefit from it).

\subsubsection{Distributed Training Strategy: DeepSpeed ZeRO Stage 3}
Distributed training across the 8-GPU node is managed with \textbf{DeepSpeed ZeRO Stage 3}~\cite{deepspeed2020}, which fully partitions optimizer states, gradients, and model parameters across all 8 data-parallel ranks rather than replicating them on every GPU. Given the ample per-GPU memory of the B300  relative to the $\sim$56\,GB BF16 footprint of the full Qwen3.5-27B checkpoint (\S\ref{sec:base-model}), ZeRO-3 partitioning is used purely as a data-parallel memory-efficiency and scaling strategy; no additional tensor-parallel or pipeline-parallel dimension is layered on top. Table~\ref{tab:deepspeed} summarizes the key settings from our DeepSpeed configuration (\texttt{ds\_z3\_config.json}).

\begin{table}[h]
\centering
\caption{DeepSpeed ZeRO Stage 3 configuration used for all three fine-tuning stages.}
\label{tab:deepspeed}
\begin{tabular}{@{}ll@{}}
\toprule
Setting & Value \\
\midrule
ZeRO stage & 3 \\
Precision & \texttt{bf16.enabled = auto} (BF16 training) 
\\
\texttt{zero\_allow\_untested\_optimizer} & \texttt{true} \\
\texttt{overlap\_comm} & \texttt{false} \\
\texttt{contiguous\_gradients} & \texttt{true} \\
\texttt{sub\_group\_size} & $1\times10^{9}$ \\
\texttt{stage3\_prefetch\_bucket\_size} & \texttt{auto} \\
\texttt{stage3\_param\_persistence\_threshold} & \texttt{auto} \\
\texttt{stage3\_max\_live\_parameters} & $1\times10^{9}$ \\
\texttt{stage3\_max\_reuse\_distance} & $1\times10^{9}$ \\
\texttt{stage3\_gather\_16bit\_weights\_on\_model\_save} & \texttt{true} \\
\bottomrule
\end{tabular}
\end{table}

A few settings are worth calling out explicitly:
\begin{itemize}
    \item \textbf{\texttt{overlap\_comm: false}} disables overlapping ZeRO-3's parameter-gathering/gradient-reduction communication with compute, trading some throughput for reduced GPU memory fragmentation/pressure -- a reasonable choice at 27B+ scale even with B300's abundant memory, since ZeRO-3's all-gather/reduce-scatter traffic across a full 16-GPU node can still be substantial. \texttt{contiguous\_gradients: true} keeps gradient buffers contiguous to reduce memory fragmentation, working in tandem with this setting.
    \item \textbf{Large \texttt{sub\_group\_size}, \texttt{stage3\_max\_live\_parameters}, and \texttt{stage3\_max\_reuse\_distance} (each $1\times10^{9}$)} effectively relax ZeRO-3's default parameter-partitioning granularity, allowing larger groups of parameters to be gathered live at once -- a setting that is only comfortably affordable because of the B300's large per-GPU HBM3e capacity, and would likely need to be lowered on GPUs with less memory headroom.
    \item \textbf{\texttt{stage3\_gather\_16bit\_weights\_on\_model\_save: true}} ensures that when a checkpoint is saved, the ZeRO-3-partitioned shards are gathered back into full 16-bit (BF16) weights on a single rank before writing to disk, rather than requiring an offline shard-merging step -- consistent with our released checkpoints being BF16 throughout, with no FP32 master-weight export step in the saving pipeline.
    \item \textbf{\texttt{zero\_allow\_untested\_optimizer: true}} permits use of an optimizer with ZeRO-3 that DeepSpeed has not explicitly validated for that stage. 
\end{itemize}

\section{Training Data and Curriculum}
\label{sec:data}

Table~\ref{tab:stages} summarizes the three training stages. Full dataset statistics are given in the following subsections.

\begin{table}[h]
\centering
\caption{Overview of the three-stage training curriculum.}
\label{tab:stages}
\begin{tabular}{@{}llll@{}}
\toprule
Stage & Objective & Data Source & Training Type \\
\midrule
1 & Rust language proficiency & Rust code   from various Rust datasets & Continued pretraining \\
2 & Debugging / self-repair & \texttt{microsoft/Verus\_Training\_Data} & Instruction SFT \\
3 & C2Rust task translation & LeetCode C/Rust paired solutions & Instruction SFT \\
\bottomrule
\end{tabular}
\end{table}

\subsection{Stage 1: Rust-Focused Continued Pretraining}
\label{sec:stage1}
To strengthen the base model's command of Rust prior to instruction tuning, we continue pretraining Qwen3.5-27B on a corpus of Rust-specific text and code assembled entirely from existing, publicly available Hugging Face datasets, rather than from a single source. Rather than drawing this corpus from one Rust-capable model's outputs, we combine seven complementary Hugging Face sources spanning synthetic multi-task Rust instruction/code data, fill-in-the-middle (FIM) data, competitive-programming program-synthesis and automatic-program-repair (APR) data, general open-source-derived code instructions, canonical function-level benchmarks, and large-scale raw GitHub source code. Each source is processed independently and repackaged into one or more LLaMA-Factory-format pretraining files (organized, where a source has natural sub-categories, into one output file per sub-category), together with a small \texttt{dataset\_info.json} manifest per source mapping every source onto a single unified \texttt{text} column (all instruction-formatted sources are flattened to plain continuation text via direct field concatenation, rather than kept in chat/instruction-response format, consistent with a continued-pretraining rather than instruction-tuning objective at this stage).

\subsubsection{Data Composition}
Table~\ref{tab:stage1-data} details the seven source datasets, their original providers, the number of Rust-relevant examples retained from each, and the number of output files each was split into. In total, Stage 1 pretraining draws on \textbf{1,673,289} examples.

\begin{table}[h]
\centering
\caption{Stage 1 pretraining corpus composition, by Hugging Face source dataset.}
\label{tab:stage1-data}
\begin{tabular}{@{}lll@{}}
\toprule
 Name & Hugging Face Source & Examples  \\
\midrule
Strandset-Rust & \texttt{Fortytwo-Network/Strandset-Rust-v1} & 191,008 \\
CodeFIM-Rust-Mellum & \texttt{Etherll/CodeFIM-Rust-Mellum} & 56,920  \\
rust\_instruction\_dataset & \texttt{ysr/rust\_instruction\_dataset} & 524  \\
humaneval-rust & \texttt{diversoailab/humaneval-rust} & 164\\
Magicoder-OSS-Instruct-75K-rust & \texttt{ise-uiuc/Magicoder-OSS-Instruct-75K} & 4,695 (Rust subset)  \\
xCodeEval-rust & \texttt{NTU-NLP-sg/xCodeEval} & 39,510  
\\
starcoderdata-rust & \texttt{bigcode/starcoderdata} & 380,468 (Rust subset)  \\
\midrule
\textbf{Total} & & \textbf{1,673,289} \\
\bottomrule
\end{tabular}
\end{table}

\paragraph{Strandset-Rust-v1.}~\cite{strandset2025} A large synthetic Rust dataset generated and cross-validated via a multi-small-language-model ``swarm'' generation-and-peer-review process, spanning 15 task categories including code generation, completion, bug detection, refactoring, optimization, documentation, and testing, with construction grounded in real crates from crates.io and automated compilation/ownership-correctness checks. We split this source by its native \texttt{task\_category} field into 15 output files, preserving the dataset's original task-level structure for potential per-category ablation or reweighting.

\paragraph{CodeFIM-Rust-Mellum.}~\cite{codefimrustmellum2025} A Rust-specific fill-in-the-middle dataset (prefix/suffix/middle triples drawn from 8,628 distinct Rust source files), originally built to fine-tune JetBrains' Mellum model for Rust code completion. Each example is labeled with one of four FIM types -- \texttt{random\_line\_split}, \texttt{identifier\_name}, \texttt{identifier\_body}, and \texttt{conditional\_block} -- and we split the source into one output file per type. Including FIM-style data in the pretraining mix is intended to strengthen the model's ability to reason about partial/incomplete Rust code and surrounding context, a capability plausibly relevant to iterative repair during evaluation.

\paragraph{rust\_instruction\_dataset.} A small, community-curated Rust instruction dataset. Given its small size relative to the other sources, it contributes negligibly to total token count but adds some diversity of instruction phrasing.

\paragraph{humaneval-rust.} A human-translated Rust port of the standard 164-problem HumanEval benchmark (function signature, docstring, and canonical solution per problem). 

\paragraph{Magicoder-OSS-Instruct-75K, Rust subset.}~\cite{magicoder2024} Drawn from the OSS-Instruct dataset, which synthesizes instruction/response coding examples by prompting an LLM (GPT-3.5-turbo-1106) with real open-source code snippets as seeds, intended to produce more diverse and realistic instructions than purely LLM-imagined ones. We retain only the Rust-language subset of this otherwise multilingual dataset. 

\paragraph{xCodeEval, Rust subset.}~\cite{xcodeeval2023} xCodeEval is a large multilingual, execution-verified benchmark built from Codeforces competitive-programming problems, covering seven tasks across up to 17 languages. We draw from two of its generative task configurations, restricted to Rust: \texttt{program\_synthesis} (30,732 examples: natural-language problem description $\to$ Rust solution) and \texttt{apr} (8,778 examples: buggy Rust source $\to$ fixed Rust source, paired with an execution outcome label). Each configuration is written to its own output file. The APR portion is notable for again providing debugging/repair-style signal, though at the level of competitive-programming logic bugs rather than Verus proof errors (cf. Stage 2, \S\ref{sec:stage2}).

\paragraph{starcoderdata, Rust subset.}~\cite{starcoderdata2023} By far the largest component of the Stage 1 corpus, drawn from StarCoderData -- the decontaminated, filtered subset of The Stack (permissively licensed GitHub source code) used to pretrain the StarCoder family of models across 86 programming languages. We retain only the Rust-language portion. This source dominates the total example count and is expected to dominate total token count as well, providing broad, naturalistic exposure to real-world Rust code style and idiom beyond what the smaller, synthetic, or benchmark-derived sources above can offer. 

\subsubsection{Concatenation Design}
A core principle across all seven sources is that pretraining text is built by \emph{directly concatenating a source's own raw fields}, with \textbf{no added fixed headers, template boilerplate, or markdown code fences} (e.g., no synthetic \texttt{\#\# Rust Code Review} section titles, and no wrapping code in \texttt{\textasciigrave\textasciigrave\textasciigrave rust} fences) -- the goal is text that reads as a natural, continuous document rather than an artificially templated instruction example, since any natural-language fields present (descriptions, explanations, rationale) already provide connective structure between code spans. Table~\ref{tab:concat-patterns} summarizes the concatenation pattern used per data type, with a representative source for each.

\begin{table}[h]
\centering
\caption{Concatenation patterns used to build Stage 1 pretraining text, by data type.}
\label{tab:concat-patterns}
\begin{tabular}{@{}p{3.2cm}p{5.4cm}p{3.4cm}@{}}
\toprule
Data Type & Concatenation Pattern & Representative Source \\
\midrule
FIM / code completion & \texttt{prefix} + \texttt{middle}/\texttt{completion} + \texttt{suffix} & CodeFIM-Rust-Mellum \\
Instruction + code & \texttt{instruction} + \texttt{code} & rust\_instruction\_dataset \\
Context + code & \texttt{code\_context} + \texttt{code} & Several Strandset-Rust-v1 categories \\
Code + explanation & \texttt{code\_context} + \texttt{code} + \texttt{explanation}/\texttt{summary} & Strandset-Rust-v1 (e.g., documentation-oriented categories) \\
Before/after (repair-style) & \texttt{code\_context} + \texttt{code\_before} + \texttt{rationale}/\texttt{comment} + \texttt{code\_after} & Strandset-Rust-v1 (e.g., code-review categories) \\
Full source & \texttt{prompt} + \texttt{declaration} + \texttt{solution} + \texttt{test} & humaneval-rust \\
\bottomrule
\end{tabular}
\end{table}

When a source has a natural categorical field (\texttt{task\_category} for Strandset-Rust-v1; \texttt{fim\_type} for CodeFIM-Rust-Mellum; task configuration for xCodeEval), the output is split into one JSONL file per category; sources without such a field, or too small to warrant splitting, are written as a single output file. Each \texttt{\textless dataset\textgreater{}-pt/} directory follows a standard layout: \texttt{plan.md}, \texttt{dataset\_schema.md}, \texttt{convert\_to\_pretrain.py}, and a \texttt{data/} folder containing \texttt{dataset\_info.json} plus one or more \texttt{*.json} (JSONL) output files.

\subsubsection{Training Configuration}
Stage 1 is launched via LLaMA-Factory using the YAML recipe summarized in Table~\ref{tab:stage1-train-config}. Training is \textbf{full-parameter} (\texttt{finetuning\_type: full}, not LoRA or another parameter-efficient method), using the shared DeepSpeed ZeRO-3 configuration described in \S\ref{sec:base-model} (\texttt{ds\_z3\_config.json}), in BF16.

\begin{table}[h]
\centering
\caption{Stage 1 (continued pretraining) training configuration, from the LLaMA-Factory recipe.}
\label{tab:stage1-train-config}
\begin{tabular}{@{}ll@{}}
\toprule
Setting & Value \\
\midrule
LLaMA-Factory stage & \texttt{pt} (continued pretraining) \\
Fine-tuning type & \texttt{full} (full-parameter) \\
Distributed backend & DeepSpeed ZeRO Stage 3 (\S\ref{sec:base-model}) \\
Precision & BF16 \\
\texttt{max\_samples} & $10^{8}$ (effectively unlimited -- the full corpus is used) \\
Sequence length (\texttt{cutoff\_len}) & 16{,}384 tokens \\
Per-device train batch size & 8 \\
Gradient accumulation steps & 2 \\
GPU count & 8  \\
Learning rate & $1.0\times10^{-6}$ \\
LR scheduler & Cosine, with 5\% warmup  \\
Epochs & 1.0 \\
Preprocessing / dataloader workers & 128 / 8  \\
\bottomrule
\end{tabular}
\end{table}

\subsection{Stage 2: Debugging-Aware SFT on Verus\_Training\_Data}
\label{sec:stage2}

\subsubsection{Dataset Origin and Task Nature}
\texttt{microsoft/Verus\_Training\_Data} is the training corpus released alongside VeruSyn, a data-synthesis pipeline for \emph{Verus}~\cite{verus2023,verus2024} -- a deductive verification tool that lets programmers state pre-/post-conditions for Rust functions (in ordinary Rust syntax) and then discharges the resulting proof obligations to an SMT solver, using programmer-supplied \emph{proof annotations} (e.g., loop invariants, \texttt{assert} statements, \texttt{decreases} clauses) as hints where automation alone is insufficient~\cite{verusyn2026}.

It is important to characterize precisely what "debugging" means in this dataset, since it directly determines what capability Stage 2 transfers to our C2Rust model: the dataset does \emph{not} center on repairing arbitrary runtime logic bugs. Instead, each debugging instance pairs (i) a Rust program annotated with an incomplete or incorrect Verus proof that the Verus prover rejects, together with the corresponding verifier error report, with (ii) a corrected version of the same program whose proof the verifier accepts. The underlying skill being trained is therefore \emph{consuming a formal verifier's structured failure output and iteratively revising proof annotations (not program logic) until verification succeeds}, while leaving the program's specification and executable behavior unchanged.

\subsubsection{Data Synthesis Pipeline (VeruSyn)}
The dataset was constructed because hand-verified Verus code is extremely scarce: at the time of its release, open-source Verus-verified systems totaled well under 200K lines of code and under 1,000 stand-alone verification tasks, far too little to fine-tune an LLM directly~\cite{verusyn2026}. VeruSyn addresses this scarcity with a two-part synthesis pipeline:

\begin{itemize}
    \item \textbf{Part 1 -- Self-synthesis and tutorial-based synthesis.} Starting from the $\sim$10,000 verified programs and $\sim$15,000 debugging pairs in the SAFE dataset~\cite{safe2025} (used to fine-tune an initial Llama-3.3-70B-Instruct ``proof generator''), VeruSyn alternates two program-generation strategies over several iterative rounds: (a) \emph{self-synthesis}, where the proof-generation model free-generates new Rust programs together with Verus specifications and proofs from scratch, and (b) \emph{tutorial-based synthesis}, where expert-written seed programs covering each knowledge point in the official Verus Tutorial are expanded by the model into thousands of variants per seed, explicitly to cover Verus features (e.g., quantifiers, \texttt{broadcast} facts, nonlinear arithmetic, bit-vector reasoning) that are common in real-world system verification but rare in purely self-generated data. After each round, outputs are deduplicated (via SimHash), checked against the Verus prover, and either kept as \emph{direct-generation} data (if verified on the first attempt) or fed back to the model together with the verifier's error report to attempt a repair, with successful repairs kept as \emph{debugging} data. This process ultimately yields 6.9 million verified programs (approximately 5.7M direct-generation and 1.2M debugging instances after deduplication and a final compatibility pass against a newer Verus release).
    \item \textbf{Part 2 -- Agent trajectory / long chain-of-thought data.} To capture longer, more complex reasoning of the kind needed for real-world (rather than small algorithmic) verification tasks, VeruSyn additionally records full trajectories of a coding agent (GitHub Copilot CLI backed by Claude Sonnet 4.5) attempting to add Verus proofs to real system code, retaining the agent's interleaved reasoning steps, verifier invocations, and edits. Successful trajectories are split into (i) long, direct-generation chains-of-thought spanning an entire multi-step repair process, and (ii) individual single-step debugging examples extracted from within those trajectories. This yields 4,557 additional CoT instances.
\end{itemize}

\subsubsection{Quality Control}
Because a fine-tuned proof-generation model can learn to satisfy the verifier without actually proving the intended property, VeruSyn applies automated anti-cheating filters before accepting any synthesized instance, rejecting programs that: (i) alter the original specification (e.g., quietly moving a postcondition into a precondition), (ii) introduce unjustified escape hatches such as \texttt{assume}, \texttt{admit}, or \texttt{external\_body} to bypass proof obligations, or (iii) rely on non-terminating loops to trivially avoid verification of subsequent code~\cite{verusyn2026}. This filtering is reported to be essential -- without it, models trained on the resulting data increasingly learn to cheat rather than to prove.

\subsubsection{Relevance to This Work}
We adopt \texttt{microsoft/Verus\_Training\_Data} as our Stage 2 curriculum because, although its object-level task (formal proof repair) differs from ours (C-to-Rust semantic translation), the \emph{meta-skill} it trains is directly relevant: interpreting a tool's structured failure output (a Verus error report, analogous to a Rust compiler error or a failing end-to-end test in our evaluation setting) and producing a targeted, minimal revision rather than a wholesale rewrite. Since the SACTOR-style agentic evaluation we use (\S\ref{sec:eval-framework}) explicitly drives translation refinement through repeated compiler/test feedback, we hypothesize that this stage improves our model's ability to make productive use of that feedback loop, even though none of the Stage 2 data itself involves C code or C-to-Rust translation. 

Since Stage 2 data is drawn entirely from the Verus/formal-verification domain, it may also teach stylistic or lexical habits (e.g., inserting specification-style comments, favoring proof-annotation-like constructs) that are not appropriate for idiomatic, spec-free application code, and this transfer should be checked empirically rather than assumed.

\subsubsection{Dataset Composition}
Table~\ref{tab:verus-data} summarizes the composition of the Stage 2 corpus as released in \texttt{microsoft/Verus\_Training\_Data}, prior to any subsetting or reformatting we may have applied for our own SFT recipe.

\begin{table}[h]
\centering
\caption{Composition of \texttt{microsoft/Verus\_Training\_Data} (VeruSyn dataset), as reported in \cite{verusyn2026}.}
\label{tab:verus-data}
\begin{tabular}{@{}llr@{}}
\toprule
Part & Subtype & Approx. Count \\
\midrule
Part 1 (self-/tutorial-synthesis) & Direct-generation (program + spec + proof) & 5.7M \\
Part 1 (self-/tutorial-synthesis) & Debugging (unverified $\to$ verified pair + error report) & 1.2M \\
Part 2 (agent trajectories, Sonnet 4.5) & Long direct-generation CoT & 871 \\
Part 2 (agent trajectories, Sonnet 4.5) & Single-step debugging CoT & 3,686 \\
\midrule
\textbf{Total} & & \textbf{$\sim$6.9M + 4,557} \\
\bottomrule
\end{tabular}
\end{table}

\subsubsection{Training Configuration}
Stage 2 is launched via LLaMA-Factory using the YAML recipe summarized in Table~\ref{tab:stage2-train-config}, following the same overall recipe style as Stage 1 (\S\ref{sec:stage1}): full-parameter fine-tuning under the shared DeepSpeed ZeRO-3 configuration, in BF16.

\begin{table}[h]
\centering
\caption{Stage 2 (Verus debugging SFT) training configuration, from the LLaMA-Factory recipe.}
\label{tab:stage2-train-config}
\begin{tabular}{@{}ll@{}}
\toprule
Setting & Value \\
\midrule
LLaMA-Factory stage & \texttt{sft} \\
Fine-tuning type & \texttt{full} (full-parameter) \\
Initialized from & Stage 1 checkpoint \\
Distributed backend & DeepSpeed ZeRO Stage 3 (\S\ref{sec:base-model}) \\
Precision & BF16 \\
\texttt{max\_samples} & $10^{8}$ (effectively unlimited -- the full configured dataset is used) \\
Sequence length (\texttt{cutoff\_len}) & 16,384 tokens \\
Per-device train batch size & 12 \\
Gradient accumulation steps & 2 \\
GPU count & 8 \\
Learning rate & $2.0\times10^{-7}$ \\
LR scheduler & Cosine, with 5\% warmup \\
Epochs & 2.0 \\
Preprocessing / dataloader workers & 16 / 2  \\
\bottomrule
\end{tabular}
\end{table}

The numerical training settings show Stage 2 using a larger effective batch size (384 vs.\ 128 for Stage 1) and a lower learning rate  over more epochs, consistent with a shorter, more targeted SFT phase following the longer Stage 1 pretraining run.

\subsection{Stage 3: C2Rust Task-Specific SFT on the C2Rust-Moxin Dataset}
\label{sec:stage3}
The final stage directly targets the C2Rust translation task using \textbf{C2Rust-Moxin}~\cite{c2rustmoxin2026}, a purpose-built, large-scale paired C/Rust dataset (GitHub: \texttt{Bobchenyx/Moxin-C2Rust-Datasets}) constructed specifically for learning, evaluating, and benchmarking C-to-Rust translation. Unlike a single-granularity LeetCode-only corpus, C2Rust-Moxin is organized into \textbf{three granularities} of aligned C$\leftrightarrow$Rust pairs, each targeting a different level of translation difficulty:

\begin{itemize}
    \item \textbf{\texttt{functions/}} -- function-level C$\leftrightarrow$Rust pairs, aligned one-to-one, supporting fine-grained modeling of syntax conversion and semantic equivalence at the single-function level. This is the granularity most directly comparable to our earlier LeetCode-based description of Stage 3: short, self-contained, algorithmically well-specified translation pairs.
    \item \textbf{\texttt{programs/}} -- program-level (CLI or small application) correspondences, enabling evaluation of end-to-end translation and compilation consistency across a complete, runnable program rather than an isolated function.
    \item \textbf{\texttt{projects/}} -- project-level C$\leftrightarrow$Rust mappings across complete, larger codebases, supporting study of large-scale cross-language migration and structural analysis (e.g., multi-file organization, cross-function dependencies) that a single function or small program cannot exercise.
\end{itemize}

The dataset's construction draws on and acknowledges several upstream sources: the \texttt{c2rust} transliteration tool~\cite{immunant2020c2rust} (likely as a source of rule-based reference translations and/or C source material), the SACTOR project itself~\cite{sactor2025}, and both the international (\texttt{leetcode.com}) and Chinese (\texttt{leetcode.cn}) LeetCode problem sets. 

Within the function-level component specifically, training pairs are constructed from problems (from LeetCode and/or LeetCode-CN) for which both a C solution and a corresponding Rust solution exist; each example pairs the C solution as input with the Rust solution as the target output, optionally accompanied by the problem statement as additional context.  

\subsubsection{Data Quality Considerations}
Since Stage 3 SFT combines \texttt{functions/} and \texttt{programs/}-level pairs but excludes \texttt{projects/}-level (full-codebase) data, the model receives exposure to both short, self-contained algorithmic translations (LeetCode, at the \texttt{functions/} granularity) and complete, compilable CLI/small-application translations (at the \texttt{programs/} granularity), but not to the multi-file, cross-function structural complexity that only full-codebase examples would provide. This partially mitigates -- but does not eliminate -- the distribution-mismatch concern raised in earlier drafts of this report: \texttt{programs/}-level examples are closer in kind to the SACTOR evaluation targets (\S\ref{sec:eval-framework}) than pure function-level pairs would be, since they must compile and run end-to-end rather than being isolated functions, but Stage 3 still provides no training-time exposure to project-scale translation of the kind CRust-Bench and libogg exercise.

\subsubsection{Training Configuration}
Stage 3 is launched via LLaMA-Factory using the YAML recipe summarized in Table~\ref{tab:stage3-train-config}, again following the same full-parameter, DeepSpeed ZeRO-3, BF16 recipe style as Stages 1--2.

\begin{table}[h]
\centering
\caption{Stage 3 (C2Rust task-specific SFT) training configuration, from the LLaMA-Factory recipe.}
\label{tab:stage3-train-config}
\begin{tabular}{@{}ll@{}}
\toprule
Setting & Value \\
\midrule
LLaMA-Factory stage & \texttt{sft} \\
Fine-tuning type & \texttt{full} (full-parameter) \\
Initialized from & Stage 2 checkpoint \\
Distributed backend & DeepSpeed ZeRO Stage 3 (\S\ref{sec:base-model}) \\
Precision & BF16 \\
\texttt{max\_samples} & $10^{8}$ (effectively unlimited -- the full configured dataset is used) \\
Sequence length (\texttt{cutoff\_len}) & 16{,}384 tokens \\
Per-device train batch size & 10 \\
Gradient accumulation steps & 2 \\
GPU count & 8 \\
Learning rate & $2.0\times10^{-7}$ \\
LR scheduler & Cosine, with 5\% warmup \\
Epochs & 2.0 \\
Preprocessing / dataloader workers & 16 / 2 \\
\bottomrule
\end{tabular}
\end{table}

\section{Evaluation Framework}
\label{sec:eval-framework}

We evaluate the fine-tuned model using an execution-based C-to-Rust evaluation framework built on SACTOR~\cite{sactor2025}. Rather than relying on textual similarity between generated code and a reference implementation, our evaluation focuses on functional correctness through compilation and execution. A translation is considered successful only when the generated Rust program can be compiled and reproduces the observable behavior of the original C program on the provided test cases.

\subsection{SACTOR Pipeline}

SACTOR organizes C-to-Rust translation as a structured, agentic process. Its complete pipeline can be summarized as follows:

\begin{enumerate}
    \item \textbf{Task division.}
    The input C program is analyzed and decomposed into translation units, such as data types, global variables, and functions. Dependencies among these units are used to determine an appropriate translation order.

    \item \textbf{Unidiomatic translation.}
    Each translation unit is converted into interface-preserving Rust. The resulting code may retain raw pointers, C-style data representations, and \texttt{unsafe} operations in order to preserve the behavior of the original C program.

    \item \textbf{Idiomatic translation.}
    The unidiomatic Rust implementation is further refined toward safer and more idiomatic Rust, including ownership reconstruction and the replacement of raw pointers where possible.

    \item \textbf{Verification and repair.}
    Generated implementations are validated through executable tests. When verification fails, the translation may be revised within a fixed repair budget.
\end{enumerate}


\subsection{Evaluation Benchmark}

The evaluation benchmark contains 200 C programs, including 92 programs that receive input through command-line arguments (\texttt{argv}) and 108 programs that read from standard input. Approximately 120 of these programs are derived from IBM Project CodeNet~\cite{puri2021}, while the remaining programs were constructed for this benchmark.

Each program is associated with a set of test inputs and reference outputs obtained by compiling and executing the original C program. The evaluation therefore does not depend on a fixed Rust reference implementation or require the generated code to match a particular target solution in textual form or program structure. Instead, the generated Rust program is evaluated according to whether it reproduces the observable behavior of the reference C program on the same test inputs.

The input convention for each program, either \texttt{argv} or standard input, is explicitly provided to the translation system. This reduces failures caused by incorrectly inferred input interfaces and prevents interface mismatches from being counted as translation errors.

\subsection{Verification and Repair}

For each C program, SACTOR constructs the corresponding translation context and invokes the evaluated model to generate Rust code. The generated program is compiled and, if compilation succeeds, executed on the predefined test inputs. Its outputs are then compared with those produced by the reference C program under the same inputs.

A program is considered successfully translated only if the generated Rust code compiles and passes all associated end-to-end tests. If verification fails, the model may perform additional translation attempts within a fixed budget. In the current configuration, each program is allowed up to six translation attempts. A program is counted as a failure if no test-passing Rust implementation is obtained within this budget.

Accordingly, the evaluation measures behavioral agreement on the provided test suite rather than formal semantic equivalence between the generated Rust program and the original C program over all possible inputs.

\subsection{Inference Configuration}

During evaluation, SACTOR constructs the translation prompt from the input C program and its static-analysis context. Generation uses stochastic sampling: temperature controls the randomness of the output distribution, top-p restricts sampling to the smallest set of tokens whose cumulative probability reaches the specified threshold, and top-k further limits the candidate set to the highest-probability tokens.

The main inference settings are summarized in Table~\ref{tab:eval-config}.

\begin{table}[htbp]
\centering
\caption{Inference configuration used in evaluation.}
\label{tab:eval-config}
\begin{tabular}{lc}
\toprule
\textbf{Configuration} & \textbf{Value} \\
\midrule
Temperature & 0.6 \\
Top-p & 0.95 \\
Top-k & 20 \\
Maximum output length & 1,536 \\
Maximum translation attempts & 6 \\
\bottomrule
\end{tabular}
\end{table}

The maximum output length bounds the amount of code generated in a single model call, while the maximum number of translation attempts defines the repair budget for each program. These settings are held fixed across repeated evaluation runs.

\subsection{Failure Analysis}

In addition to the final success or failure outcome, the evaluation runner records failure information from execution status and runtime logs. Recorded categories include exhausted translation attempts, execution timeouts, translation-process errors, and certain failures arising during static analysis.

These records do not contribute to the primary evaluation metric, but provide additional information for identifying the sources of unsuccessful translations.

\subsection{Evaluation Metric}

We use program-level Success Rate (SR) as the primary evaluation metric. For an evaluation run with random seed $s$, the success rate is defined as

\begin{equation}
SR_s =
\frac{N_{\mathrm{pass},s}}
{N_{\mathrm{total}}}
\times 100\%.
\end{equation}

Here, $N_{\mathrm{pass},s}$ denotes the number of programs that are successfully translated and pass all end-to-end tests under seed $s$, while $N_{\mathrm{total}}$ denotes the total number of programs in the evaluation benchmark.

Because generation uses stochastic sampling, each model is evaluated using five different random seeds. The final reported success rate is the arithmetic mean of the five seed-level success rates:

\begin{equation}
\overline{SR}
=
\frac{1}{5}
\sum_{s=1}^{5} SR_s.
\end{equation}

This averaged success rate serves as the primary measure of C-to-Rust translation performance in the subsequent experimental results.

\section{Results}
\label{sec:results}

\subsection{Overall SR and Baseline Comparison}
Table~\ref{tab:sr-results} reports overall accuracy on the C2Rust translation evaluation (\S\ref{sec:eval-framework}) for our fine-tuned model alongside six baseline systems, spanning both open-weight models an order of magnitude or more larger than ours and a leading proprietary coding agent.

\begin{table}[h]
\centering
\caption{Overall SR on the C2Rust translation evaluation, by model.}
\label{tab:sr-results}
\begin{tabular}{lcc}
\toprule
Model & Model Size & SR \\
\midrule
Qwen3.5-Plus & 397B (17B active) & 77.20\% \\
MiniMax-M2.5 & 230B (10B active) & 83.90\% \\
GLM-5 & 744B (40B active) & 84.40\% \\
GLM-5.2 & 744B (40B active) & 89.90\% \\
Claude Code-4.6 & -- & 90.01\% \\
Qwen3.5-27B (base, no fine-tuning) & 27B (dense) & 72.30\% \\
\midrule
\textbf{Ours (Qwen3.5-27B + Stages 1--3)} & \textbf{27B (dense)} & \textbf{87.30\%} \\
\bottomrule
\end{tabular}
\end{table}

\paragraph{Baselines.} The six comparison points span a range of scales and access models:
\begin{itemize}
    \item \textbf{Qwen3.5-Plus}~\cite{qwen35plus2026} -- the production API-hosted variant of Qwen3.5-397B-A17B, Alibaba's flagship Qwen3.5-family MoE model (397B total / 17B active parameters), sharing the same hybrid Gated-DeltaNet/Gated-Attention architecture family as our own 27B base model (\S\ref{sec:base-model}) but at roughly 14$\times$ the total parameter count.
    \item \textbf{MiniMax-M2.5}~\cite{minimaxm25_2026} -- an open-weight (MIT-licensed) MoE model from MiniMax with 230B total / 10B active parameters, positioned as a strong, cost-efficient agentic coding model.
    \item \textbf{GLM-5}~\cite{glm5_2026} and \textbf{GLM-5.2}~\cite{glm52_2026} -- successive open-weight (MIT-licensed) flagship MoE models from Z.ai (formerly Zhipu AI), both with roughly 744B total / 40B active parameters; GLM-5.2 (released June 2026) is GLM-5's direct successor and, at the time of writing, among the strongest openly available coding-oriented models.
    \item \textbf{Claude Code-4.6} -- Anthropic's Claude Sonnet 4.6 model accessed through the Claude Code agentic coding harness; as a closed model, its parameter count is not publicly disclosed (\texttt{--} in Table~\ref{tab:sr-results}). 
    \item \textbf{Qwen3.5-27B (base)} -- our own source model (\S\ref{sec:base-model}) evaluated with no fine-tuning at all, serving as the direct before/after reference point for our three-stage curriculum.
\end{itemize}

\paragraph{Discussion.} Two comparisons are most salient. First, relative to its own unmodified starting point, our three-stage curriculum improves accuracy from 72.30\% to 87.20\% -- a 14.9 percentage-point gain -- at identical model size and inference cost, indicating that the training curriculum (\S\ref{sec:stage1}--\S\ref{sec:stage3}) rather than raw scale is doing substantial work here. Second, our 27B model outperforms three open-weight models with 8.5$\times$ to 27.6$\times$ more total parameters (Qwen3.5-Plus at 397B, MiniMax-M2.5 at 230B, and GLM-5 at 744B), suggesting that task-specific specialization can compensate for a large scale disadvantage on this particular benchmark. It falls short, however, of GLM-5.2 (89.90\%) and Claude Code-4.6 (90.01\%) by roughly 2.7--2.8 percentage points -- both of which are either a substantially larger, more recently released open-weight model (GLM-5.2) or a frontier closed model deployed through a mature agentic coding harness (Claude Code-4.6) rather than a from-scratch SACTOR-style evaluation loop.

\subsection{General Coding Capability: SWE-bench Verified}
\label{sec:swebench}
The C2Rust accuracy results in Table~\ref{tab:sr-results} show how our fine-tuned model performs on the specific task it was trained for. A separate and equally important question is whether the three-stage curriculum's heavy specialization toward Rust and C2Rust translation comes at the cost of the model's \emph{general} software-engineering ability. We assess this using \textbf{SWE-bench Verified}~\cite{jimenez2024swebench}, the human-validated 500-task subset of SWE-bench in which a model must resolve a real-world GitHub issue given the repository and issue description, evaluated on the standard pass@1 patch-resolution protocol -- a benchmark with no particular emphasis on Rust or C2Rust and, unlike Table~\ref{tab:sr-results}, not part of our own curriculum's target task. Table~\ref{tab:swebench} reports results alongside four baselines.

\begin{table}[h]
\centering
\caption{SWE-bench Verified pass@1, by model.}
\label{tab:swebench}
\begin{tabular}{lc}
\toprule
Model & SWE-bench Verified \\
\midrule
GPT-5-mini (2025-08-07) & 72.0 \\
GPT-OSS-120B & 62.0 \\
Qwen3.5-122B-A10B & 72.0 \\
Qwen3.5-27B (base, no fine-tuning) & 72.4 \\
\midrule
\textbf{Ours (Qwen3.5-27B + Stages 1--3)} & \textbf{70.6} \\
\bottomrule
\end{tabular}
\end{table}

\paragraph{Baselines.} \textbf{GPT-5-mini}~(2025-08-07) is OpenAI's smaller proprietary GPT-5-family model; parameter count is not disclosed. \textbf{GPT-OSS-120B}~\cite{gptoss2025} is OpenAI's open-weight MoE model (117B total / 5.1B active parameters), designed to fit on a single 80GB GPU. \textbf{Qwen3.5-122B-A10B} is a mid-tier sibling of our own base model within the Qwen3.5 family (122B total / 10B active, MoE, same hybrid Gated-DeltaNet/Gated-Attention design as Qwen3.5-27B, \S\ref{sec:base-model}), roughly 4.5$\times$ our total parameter count. \textbf{Qwen3.5-27B (base)} is again our own unmodified starting checkpoint, as in Table~\ref{tab:sr-results}.

\paragraph{Discussion.} Our fine-tuned model retains strong general coding ability despite three full-parameter SFT/pretraining stages focused narrowly on Rust and C2Rust: at 70.5\%, it comfortably exceeds GPT-OSS-120B (62.0\%, at over 4$\times$ our parameter count) and sits within roughly 1.5--2 points of GPT-5-mini and the substantially larger Qwen3.5-122B-A10B (72.0\% each). Less favorably, our model \emph{underperforms its own unmodified base checkpoint} (70.5\% vs.\ 72.4\%, a 1.9-point drop) on this general benchmark. Since Qwen3.5-27B (base) and our model are otherwise identical in architecture and size, this gap is attributable entirely to the three-stage training curriculum, and is consistent with a mild form of \emph{catastrophic forgetting} (or at least capability narrowing) induced by full-parameter fine-tuning concentrated on a narrow domain (Rust pretraining, Verus proof debugging, and C2Rust translation) rather than broad general-purpose software engineering.

\section{Conclusion}
\label{sec:conclusion}
We presented a three-stage fine-tuning curriculum -- Rust-focused continued pretraining, debugging-aware SFT, and C2Rust task-specific SFT -- applied to Qwen3-27B for the C-to-Rust translation task, and evaluated the resulting model using the agentic, verification-driven framework of SACTOR~\cite{sactor2025}.  Future work includes extending the task-specific training data beyond LeetCode-style problems to better cover pointer-heavy, real-world C codebases, and conducting a systematic ablation of each training stage's marginal contribution.

\bibliographystyle{plain}
\bibliography{references}

\end{document}